\documentclass[twocolumn,showpacs,preprintnumbers,amsmath,amssymb]{revtex4}
\usepackage{epsfig}

\usepackage{graphicx}
\usepackage{amsmath,amssymb}
\usepackage{natbib}
\newcommand\bea{\begin{eqnarray}}
\newcommand\eea{\end{eqnarray}}
\newcommand\beq{\begin{equation}}
\newcommand\eeq{\end{equation}}

\def\nn{\nonumber}
\def\dg{\dagger}
\def\f{\frac}
\def\la{\langle}
\def\ra{\rangle}

\def\b{\beta}

\def\e{\epsilon}

\def\bk{{\bf k}}

\def\bx{{\bf x}}
\def\b0{{\bf 0}}

\def\nn{\nonumber}

\begin{document}

\title{Symmetric and asymmetric charge transport in interacting asymmetric quantum impurities}
\author{Dibyendu Roy}
\affiliation{$^1$Department of Physics, University of California-San Diego, La Jolla, California 92093-0319}
%\date{\today} 

\begin{abstract}
We study steady-state charge transfer across an interacting resonance-level model connected asymmetrically to two leads. For a linear energy dispersion relation of the leads, we calculate current-voltage characteristics of the model exactly employing the scattering Bethe-Ansatz of Mehta-Andrei and find symmetric transport showing the absence of diode effect. Next we study a lattice version of this model with a nonlinear dispersion for the leads using the Lippmann-Schwinger scattering theory. We find that the inclusion of nonlinearity in the leads' dispersion causes rectification for asymmetric junctions but does not rectify for asymmetric interactions and perfect junctions. The model in the latter case can be mapped into a model of a single noninteracting electron in higher dimensions.
\end{abstract}      

\maketitle
\section{Introduction}
Rectification is considered as current asymmetry for the forward and the reverse bias. In the past years, rectification in nanoscale coherent systems has got a lot of interest. Current rectification by single asymmetric organic molecules has been predicted \cite{Aviram74} and realized experimentally \cite{Geddes90, Zhao05}. There are also several theoretical and experimental studies of charge \cite{Song98, Stopa02} and spin \cite{Ono02} rectification in different mesoscopic semiconductor heterostructures. The future application of modern molecular electronics largely depends on the high-quality molecular rectifiers. Thus one needs to understand the basic mechanism of rectification by molecules, i.e., what are the necessary and sufficient conditions such that molecular junctions act as rectifier? The mechanism of rectification by molecules is highly debated \cite{Datta97,Mujica02,Kornilovitch02, Stokbro03}.

The mechanism of current rectification in the original semiconductor $p$-$n$ junction diodes or the Schottky diodes consisting of metal-semiconductor junction is a mismatch of band structures which creates a potential barrier that blocks the motion of carriers in one direction while allowing them to flow in the opposite direction. In fact, there are also microscopic studies in the recent past along this direction with different hybrid structures showing charge and thermal energy rectification. But one finds current asymmetry in molecular junctions or nanostructures even for similar types of electrodes. Spatial asymmetry and nonlinear interaction between carriers are regarded as the necessary conditions for charge rectification in these systems but it is still not clear what are the sufficient conditions for rectification \cite{Wu09}. Quantum impurities are the simplest models for molecules or nanoscale heterostructures. Here we examine charge transport in a quantum impurity, namely, interacting resonance-level model (IRLM) connected asymmetrically to two leads. The equilibrium physics of the IRLM is well studied \cite{Filyov80}, and recently the nonequilibrium transport in the IRLM has been received a lot of interest \cite{Mehta06, Doyon07, Borda08, Boulat08}. We employ linear and nonlinear energy dispersion of the leads. Surprisingly we find symmetric charge transport in the IRLM for linear dispersion of the leads even with different tunneling junctions within the scattering Bethe-Ansatz approach of Mehta and Andrei (MA) \cite{Mehta06}. The inclusion of nonlinearity in the leads' dispersion causes rectification for asymmetric junctions but does not cause rectification for asymmetric interactions and perfectly transmitting junctions. Our model in the latter case can be viewed as a single noninteracting electron in the presence of elastic barriers in higher dimensions. %We try to understand our results physically and find sufficient conditions for charge rectification by quantum impurities. 

Rectification in asymmetric impurities is essentially a nonlinear transport phenomenon. Study of charge transfer across out-of-equilibrium quantum impurities  has attracted much attention theoretically \cite{Wingreen94, Schiller95, Mehta06,Doyon07, Dhar08, Borda08,Boulat08, RoySoori09} as well as experimentally \cite{Ralph94} for quite some time. The nonequilibrium steady-state properties of quantum impurities can be investigated  within the time-independent scattering approach. Recently MA \cite{Mehta06} have developed a nonperturbative framework generalizing the equilibrium Bethe-Ansatz to compute steady-state properties of an IRLM connected symmetrically to left and right leads with a finite chemical-potential difference. They have employed a linear energy dispersion relation for the leads which is necessary for the application of the Bethe-Ansatz in their technique. Here first we apply the scattering Bethe ansatz framework to  derive an exact expression for  the charge current through the IRLM connected asymmetrically to the two leads with linear energy dispersion. For nonlinear dispersion of the leads, one can use different theoretical techniques such as the nonequilibrium Green's function formalism \cite{Wingreen94} or the $ab~initio$ first-principles calculations \cite{Stokbro03}. One needs to make many approximations to apply these techniques for any interacting model, thus practically it is not possible to derive nonlinear transport in lattice models exactly. Recently we have studied nonequilibrium charge transport in quantum impurities with sinusoidal dispersion of the leads using the Lippmann-Schwinger (LS) scattering theory \cite{Dhar08, RoySoori09}. We here employ that method to study charge transfer in a lattice version of the IRLM with nonlinear dispersion of the leads and asymmetric junctions or interactions. Our purpose in this paper is also to facilitate a critical discussion on the above two methods investigating nonlinear transport in interacting quantum impurities. 

\section{Scattering Bethe-Ansatz for linear dispersion} The IRLM consists of a resonant level of energy $\epsilon_d$  connected to two leads via tunneling junctions of strengths $t_1$ and $t_2$ and Coulomb interaction  $U$ between the level and the leads. Then we apply standard manipulations for impurity models where we keep only the $s$ angular modes around the impurity and linearize the bath spectrum around the Fermi energies. The Hamiltonian of the system as chiral $1-d$ field theories is given by
\bea 
\mathcal{H}&=&-i \sum_{\alpha=1,2}\int dx~ \psi^\dg_{\alpha}(x)\partial \psi_{\alpha}(x) + \epsilon_d d^\dg d + \f{1}{\sqrt{2}}\big(t_1\psi^\dg_{1}(0)d \nn \\
&+&t_2\psi^\dg_{2}(0)d+ H.c.\big)+U\sum_{\alpha=1,2}\psi^\dg_{\alpha}(0)\psi_{\alpha}(0)d^\dg d~.
\label{HamAsy}
\eea
where we need to introduce a cut-off (bandwidth $D$) to make the model finite in the renormalized theory. Also we need to take same Fermi velocity for the both leads which we set to unity here. The current operator in this model is defined as $I=i\big(t_1\psi^\dg_{1}(0)d-t_2\psi^\dg_{2}(0)d-H.c.\big)/(2\sqrt{2})$. 
%\bea
%I=\f{i}{2\sqrt{2}}\big(t_1\psi^\dg_{1}(0)d-t_2\psi^\dg_{2}(0)d-H.c.\big) \nn
%\eea
Once we  compute the many-particle scattering eigenstate $|\Psi\ra_s$ for the asymmetric model, we can determine the steady-state current between the two leads by taking expectation of $I$ in $|\Psi\ra_s$  
\bea
\la I \ra =\f{\la \Psi|I|\Psi\ra_s}{\la \Psi|\Psi\ra_s} \label{currexp}
\eea
Now we transform the field operators $\psi_{1}(x),\psi_{2}(x)$ of the two leads to a new set of even and odd field operators $\psi_{e}(x), \psi_{o}(x)$ using $\psi_{1}(x)=(t_1\psi_{e}(x)+t_2\psi_{o}(x))/\sqrt{t_1^2+t_2^2}$ and $\psi_{2}(x)=(t_2\psi_{e}(x)-t_1\psi_{o}(x))/\sqrt{t_1^2+t_2^2}$~.
%\begin{eqnarray}
%\left(\begin{array}{l} \psi_{1}(x) \\ \psi_{2}(x)\end{array}
%\right)=\f{1}{\sqrt{t_1^2+t_2^2}}\left(
%\begin{array}{ll}
% t_1 &~~~~~ t_2\\
%t_2 &~~ -t_1 \end{array}\right)\left(\begin{array}{l} \psi_{e}(x) \\ \psi_{o}(x)\end{array}
%\right)
%\label{Smat}
%\end{eqnarray}
Under the transformation, the Hamiltonian in Eq.(\ref{HamAsy}) is decomposed into two parts of the even and the odd field operators.
\bea 
\mathcal{H}&=&\mathcal{H}_e+\mathcal{H}_o ~~~~~{\rm with}\label{HamAsy1} \\
\mathcal{H}_e &=&-i \int dx~ \psi^\dg_{e}(x)\partial \psi_{e}(x) + t \big(\psi^\dg_{e}(0)d+d^\dg\psi_{e}(0)\big) \nn \\  
&& + \epsilon_d d^\dg d +U \psi^\dg_{e}(0)\psi_{e}(0)d^\dg d~,\nn \\
\mathcal{H}_o &=&-i \int dx~ \psi^\dg_{o}(x)\partial \psi_{o}(x) +U \psi^\dg_{o}(0)\psi_{o}(0)d^\dg d~,\nn
\eea
with $t=\sqrt{(t_1^2+t_2^2)/2}$. Then the Hamiltonian in Eq.(\ref{HamAsy1}) is exactly similar in the form of the Hamiltonian obtained by MA in their symmetric IRLM study after  the transformation to a symmetric/antisymmetric basis. Thus the steps from here to evaluate the current are quite similar to that of Ref.\cite{Mehta06}. 
First we calculate single-particle scattering states for the different boundary conditions (i.e., incoming electron from different leads) by solving the single-particle Schr{\"o}dinger equation (here we have incorporated discontinuities at $x=0$ following MA). We define the single-particle scattering states $|1,p\ra$ for those with an incoming particle from lead 1.
\bea
|1,p\ra =&\int& dx ~ e^{ipx} \Big [ \f{1}{1+e^{i\delta_p}} \big ( [2\theta (-x)+\f{t_2^2+t_1^2e^{i\delta_p}}{t^2}]\psi_1^{\dg}(x)\nn \\
&+&(e^{i\delta_p}-1)\f{t_1t_2}{t^2}\theta(x)\psi_2^{\dg}(x)\big)+\f{e_pt_1}{\sqrt{2}t}\delta (x)d^{\dg}\Big ]|0\ra \nn
\eea
where $\delta_p=2 \arctan [t^2/2(p-\epsilon_d)]$ and $e_p=t/(p-\epsilon_d)$. We get the state $|2,p\ra$ (those with an incoming particle from lead 2) from the state $|1,p\ra$ by interchanging simultaneously the field operators $\psi_1^{\dg}(x)$ and $\psi_2^{\dg}(x)$ as well as the tunnelings $t_1$ and $t_2$.
The many-particle scattering state is constructed from the single-particle scattering eigenstates using the open Bethe-Ansatz framework \cite{Mehta06}.  For that we have to calculate two-particle $S$ matrix by finding the two-particle scattering states  for different boundary conditions. The linear dispersion in the leads gives the freedom to choose the two-particle $S$ matrix between all electrons  to be the same; this helps to   generalize the construction to many particle scattering state $|\Psi\ra_s$.  Next one forms a Bethe-Ansatz basis of eigenstates for the noninteracting electrons in the leads. The nonequilibrium boundary condition (namely, the different chemical potentials of the leads) has been incorporated in incoming particles' Bethe-Ansatz momenta $\{p_j\}$ in $|\Psi\ra_s$ which are determined by solving the Bethe-Ansatz equations. 
 
Now following the above stated prescriptions, we evaluate the many-particle scattering state $|\Psi\ra_s$ and the corresponding Bethe-Ansatz momenta. Then we use Eq.(\ref{currexp}) to find the steady-state current between the leads. Finally taking the thermodynamic limit at zero temperature, we get 
\bea
\la I \ra_s=\int dp [\rho_1(p)-\rho_2(p)]\f{\Gamma_1\Gamma_2}{(p-\epsilon_d)^2+(\Gamma_1+\Gamma_2)^2/4}
\label{current}
\eea
with $\Gamma_{\alpha}=t_{\alpha}^2/2$, where the distribution functions $\rho_{\alpha}(p)~(\alpha=1,2)$ satisfy following coupled equations \cite{Mehta06, Mehta07},
\bea
\rho_{\alpha}(p)&=&\f{1}{2\pi}\theta (k_0^{\alpha}-p)-\sum_{\beta=1,2}\int_{-\infty}^{k^{\beta}_0}K(p,k)\rho_{\beta}(k)dk~ \label{dist}\\
{\rm with}&&K(p,k)=\f{U}{\pi}\f{(\epsilon_d-k)}{(p+k-2\epsilon_d)^2+\f{U^2}{4}(p-k)^2}.\nn
\eea
The Bethe momenta $p$ in the lead $\alpha$ are filled from the lower cut-off $(-D)$ up to $k_0^{\alpha}$ which is derived from,
\bea
\int_{-D}^{\mu_{\alpha}}\f{1}{2\pi}dp=\int_{-D}^{k_0^{\alpha}}\rho_{\alpha}(p) dp \nn
\eea 
Equation (\ref{current}) correctly reproduces the result of Refs. \cite{Mehta06} and \cite{Mehta07} for the symmetric tunneling junctions, i.e., $t_1=t_2$. One can solve the above coupled equations to find the nonequilibrium distribution of the Bethe momenta using Wiener-Hopf method for $U \to \infty$ or numerically for arbitrary $U$. As the Eq.(\ref{dist}) determining the distribution functions of the Bethe momenta are independent of the tunneling junctions $t_1$ and $t_2$, we find from Eq.(\ref{current}) that  current between the leads remains the same if we interchange the tunneling junctions between the left and the right leads keeping the chemical potential of the leads fixed. Thus the transport is symmetric for the forward and the reverse bias, i.e., for the chemical potentials $[\mu+V,\mu]$ and $[\mu,\mu+V]$, where $V$ is the bias (we have set charge as unity everywhere). It shows that there is no diode effect or rectification in the current-voltage characteristics of this model even in the presence of the spatial asymmetry and the nonlinear interaction between electrons.

\section{Lippmann-Schwinger scattering theory for nonlinear dispersion} At this point the obvious question comes to our mind; what happens with the inclusion of nonlinearity in the dispersion relation of the leads?  As we have discussed before  we study the transport problem for nonlinear dispersion using the LS scattering approach \cite{Dhar08, RoySoori09}. We here consider a lattice version of the IRLM; the Hamiltonian is given by
\bea
H&=&H_0+V~, \label{disIRLM} \\
{\rm where}~~H_0&=&-\sum_{x=-\infty}^{\infty}\hspace{-0.15cm}'(c^{\dag}_xc_{x+1}+c^{\dag}_{x+1}c_x)+\epsilon_dn_0 \nn \\
&&-(t_1 c^{\dag}_{-1}c_{0}+ t_2 c^{\dag}_{0}c_{1} +{\rm H.c.})~,\nn \\
{\rm and}~~V&=&U_1n_{-1}n_0+U_2n_0n_1~, \nn
\eea
where $n_x=c^{\dag}_xc_x$ is the number operator at site $x$. $\sum '$ denotes omission of $x=-1,0$ from the summation. The Hamiltonian in Eq.(\ref{disIRLM}) describes a resonant level of energy $\epsilon_d$ at site $0$ being coupled with two noninteracting leads of spinless electrons modeled by one-dimensional tight-binding lattice. We set the lattice spacing and $\hbar$ to 1. Also we have taken  an arbitrary strength for the tunneling junctions as well as the interaction between the resonant level and the left/right leads. 

The energy dispersion of a single particle with wave number $k$ is given by $E_k = -2 \cos k$, where $- \pi < k < \pi$. The wave function $\phi_k (x)$ for a 
particle incident on the resonant level from the left (with $0 < k < \pi$) or from the 
right (with $-\pi < k < 0$) can be found in terms of the tunneling $t_1,t_2$ 
and $\epsilon_d$. 
The transmission probability $|t_k|^2$ turns out to be the same for wave numbers $k$ and $-k$. We will find later that the two-particle current may not have 
 this symmetry as a result of the interactions. We can determine the two-particle energy eigenstate for this model exactly \cite{Dhar08, RoySoori09}. The noninteracting two-particle energies and wave functions of $H_0$ are given by $E_{\bk}=E_{k_1}+E_{k_2}$ and $\phi_{\bk} (\bx) = \phi_{k_1} (x_1)\phi_{k_2} (x_2) - \phi_{k_1} (x_2) \phi_{k_2} (x_1)$, where $\bk=(k_1,k_2)$ and $\bx=(x_1,x_2)$. A scattering eigenstate (in the position basis) of the total Hamiltonian $H $ is given by the LS equation
\bea
\psi_{\bk}(\bx)&=&\phi_{\bk}(\bx) + U_1 K'_{E_{\bk}}(\bx)\psi_{\bk}(-1,0)\nn \\
&+&U_2 K_{E_{\bk}}(\bx)\psi_{\bk}(0,1) \label{scatt}
\eea
where $K'_{E_{\bk}}(\bx)=\la \bx|G_0^{+}(E_{\bk})|-1,0\ra$ and $K_{E_{\bk}}(\bx)=\la \bx|G_0^{+}(E_{\bk})|0,1\ra$ with $G_0^+(E_{\bk}) = {1}/{(E_{\bk}- H_0 +i \e)}$.
  The subscript ${\bf k}$ in the full scattering state $\psi_{\bk}(\bx)$ in Eq.(\ref{scatt}) denotes the momenta of the incoming electrons. The momenta $\{k'_1,k'_2\}$ of the scattered electrons can be quite different from the incident momenta $\{k_1,k_2\}$, but must satisfy the total energy conservation after elastic scattering which is given by $\cos k_1+\cos k_2=\cos k'_1 + \cos k'_2$. Now it is easy to prove the following properties from the above definitions, $K'_{E_{\bk}}(-1,0)=K_{E_{\bk}}(0,1)=K_0$ (say) and $K'_{E_{\bk}}(0,1)=K_{E_{\bk}}(-1,0)=K_1$ (say).  Then $\psi_{\bk}(-1,0)$ and $\psi_{\bk}(0,1)$ in Eq.(\ref{scatt}) are found in terms of these matrix elements.
\bea
\psi_{\bk}(-1,0)&=&\f{K_1U_2\phi_{\bk}(0,1)+(1-U_2K_0)\phi_{\bk}(-1,0)}{1-(U_1+U_2)K_0+U_1U_2(K_0^2-K_1^2)} \nn \\
\psi_{\bk}(0,1)&=&\f{K_1U_1\phi_{\bk}(-1,0)+(1-U_1K_0)\phi_{\bk}(0,1)}{1-(U_1+U_2)K_0+U_1U_2(K_0^2-K_1^2)} \nn 
\eea
Thus the two-particle scattering states are determined fully using the above $\psi_{\bk}(-1,0)$ and $\psi_{\bk}(0,1)$ in Eq.(\ref{scatt}). The many-particle scattering states for the nonlinear dispersion of the leads can be calculated within a two-particle scattering approximation. This is surely a perturbative approach for many particles, but it can be justified for weak interaction and/or weak tunneling with lower density of electrons in the leads.    

\begin{figure}[t]
\begin{center}
\includegraphics[width=8.5cm]{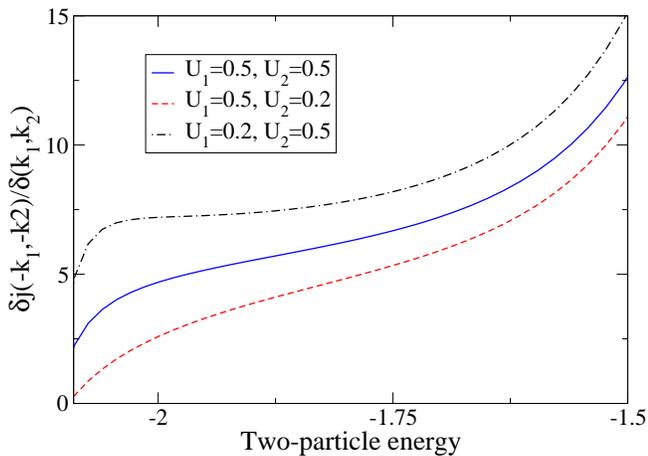}
\end{center}
\caption{(Color online) Plot of the ratio $\delta j(-k_1,-k_2)/\delta j(k_1,k_2)$ versus the energy of  two incident electrons for $t_1=0.2,t_2=0.5$ and $\epsilon_d=-0.5$. $|k_2|$ is kept fixed at $1.2$ while $|k_1|$ is changed from $0.8$ to $1.15$~.}
\label{rectification}
\end{figure}

Now we calculate the steady-state current in this model at zero temperature. The current operator on the leads is defined by, $j_x=-i(c^{\dg}_xc_{x+1}-c^{\dg}_{x+1}c_x)$. First we find two-particle current $j(k_1,k_2)$ by taking expectation value of $j_x$ in the two-particle scattering state $|\psi_{\bk}\ra=|\phi_{\bk}\ra+|S_{\bk}\ra$ [from Eq.(\ref{scatt})]. $j(k_1,k_2)=j_I+j_C+j_S$, where current in the incident state is  $j_I=\la \phi_\bk | j_x |\phi_\bk \ra= 2 \mathcal{N} (\sin k_1 |t_{k_1}|^2
+ \sin k_2 |t_{k_2}|^2)$, and the contribution from the scattered wave functions are $j_C = \la \phi_\bk | j_x | S_\bk \ra + \la S_\bk | j_x | \phi_\bk \ra$, and $j_S = \la S_\bk | j_x | S_\bk \ra$. The normalization factor $\cal N$ in $j_I$ will disappear in the many-particle current. We calculate the change in two-particle current, $\delta j(k_1,k_2)=j_C+j_S$ numerically. 
We find that $\delta j(k_1,k_2) \ne -\delta j(-k_1,-k_2)$ if $t_1 \ne t_2$ even for $U_1=U_2$. This implies that the two-particle current change due to the interaction is quite different for the particles incident from the left as compare to the right even at the same energy. 
%This seems to be a violation of time-reversal symmetry though it is not an actual violation here. 
We have seen similar asymmetry in \cite{RoySoori09} for another impurity model. This asymmetry in the two-particle current in the presence of interactions and asymmetric junctions is the reason for rectification in the many-particle current, and the amount of rectification will be larger with increasing two-particle current asymmetry. So in Fig.(\ref{rectification}), we plot the ratio $\delta j(-k_1,-k_2)/\delta j(k_1,k_2)$ versus the energy of  two incident electrons for the asymmetric junctions and different values of $U_1$ and $U_2$. There is a large asymmetry in the two-particle current for a value of $\epsilon_d$ corresponding to a two-particle resonance \cite{RoySoori09}. Now we evaluate the many-particle current, $j=j_I+\delta j$. First we take the chemical potential of the left and the right lead respectively $\mu_{L} = - 2 \cos (k_0+\delta k)$ and $\mu_{R} = - 2 \cos (k_0)$ with bias $V=\mu_{L}-\mu_R$. Here $k_0+\delta k$ $(-k_0)$ is the highest occupied wave number of the left (right) lead. In the thermodynamic limit, the noninteracting current, $j_I=\int_{-k_0}^{k_0+\delta k} (dk/2\pi) 2 \sin k |t_k|^2$ remains the same in magnitude when we reverse the bias. Thus we just need to find the many-particle current change $\delta j$ due to interactions to see rectification. Within the two-particle scattering approximation, $\delta j$ is given by
\bea 
\delta j =\int_{k_0}^{k_0+\delta k} \Big[ \int_{-k_0}^{k_0} + \f{1}{2} 
 \int_{k_0}^{k_0+\delta k} \Bigr] \f{dk_1 d k_2}{(2 \pi)^2} \delta j(k_1, k_2). 
\label{curr} 
\eea
We can write a similar expression for the $\delta j$ with the reverse bias, i.e., $k_0$ $(-[k_0+\delta k])$ is the highest occupied wave number of the left (right) lead. It can be checked that the many-particle current change will be asymmetric, i.e., $\delta j (V) \ne -\delta j(-V)$, in the presence of the two-particle current asymmetry by inspecting the $\delta j$ for the forward and the reverse bias. Thus we find here that an interacting quantum impurity acts as a rectifier for asymmetric coupling to the leads with nonlinear energy dispersion. 

Finally we simplify the lattice model in Eq.(\ref{disIRLM}) by considering $t_1=t_2=1$ and $\epsilon_d=0$ which corresponds to a perfectly transmitting impurity in the absence of interaction. Now the single-particle state $\phi_{\bk}(x)=e^{ikx}$ and $K_0$, $K_1$ are explicitly given by two-dimensional lattice Green's function $g^{+}_{E_{\bk}}(\bx)$. $K_0=g^{+}_{E_{\bk}}(0,0)-g^{+}_{E_{\bk}}(-1,1)$ and $K_1=g^{+}_{E_{\bk}}(-1,-1)-g^{+}_{E_{\bk}}(-2,0)$ where $g^{+}_{E_{\bk}}(\bx)$ can be found in terms of the complete elliptic integrals \cite{Morita71}. The two-particle current change in this simplified model has been derived analytically for arbitrary values of the $U_1$ and $U_2$. Though the total current change $\delta j(k_1,k_2)$ is same  for the left/right lead but $j_C$ and $j_S$ separately are different in the different leads. For $k_1,k_2>0$, $j_C(x>0)=2 \Upsilon$ and $j_C(x<0)=0$; for $k_1,k_2<0$, $j_C(x>0)=0$ and $j_C(x<0)=-2 \Upsilon$; and for $k_1,k_2$ opposite signs, $j_C(x>0)=\Upsilon$, and $j_C(x<0)=-\Upsilon$, where    
\bea
\Upsilon=2~{\rm Im}[U_1\psi_{\bk}(-1,0)\phi^{*}_{\bk}(-1,0)+U_2\psi_{\bk}(0,1)\phi^{*}_{\bk}(0,1)]~. 
\eea
Similarly we find for $k_1,k_2 \gtrless 0$
\bea
&&j_S(x>0)=-2\big[U^2_1|\psi_{\bk}(-1,0)|^2+U^2_2|\psi_{\bk}(0,1)|^2\big]{\rm Im}[K_0]\nn \\
&&+U_1U_2~{\rm Im}\Big[\int_{-\pi}^{\pi}dq_1\Big(\f{\Lambda(2e^{iq_1+2iQ}-e^{3iQ}-e^{2iq_1+iQ})}{4\pi\sin^2Q} \nn \\&&+\f{\Lambda^*(2e^{-iq_1}-e^{-iQ}-e^{-2iq_1+iQ})}{4\pi\sin^2Q}\Big)\Big]~,\label{js}
\eea 
where $\Lambda= \psi_{\bk}(-1,0)\psi^*_{\bk}(0,1)$ and $E_{\bk}-E_{q_1}=-2\cos Q$. While $j_S(x<0)$ is given by Eq.(\ref{js}) with $Q$ being replaced by $-Q$ and  the sign of the coefficient of Im$[K_0]$ being positive. Interestingly now we find $\delta j(k_1,k_2)=-\delta j(-k_1,-k_2)$ for arbitrary values of $U_1$ and $U_2$. Also $\delta j(k_1,k_2)=0$ for opposite signs of $k_1$ and $k_2$. So there is no rectification in the simplified version within the two-particle scattering approximation. The simplified model with two electrons can be mapped into a model of a noninteracting electron in two dimensions with two impurity sites of strength $U_1$ and $U_2$ at $(-1,0)$ and $(0,1)$. Similarly for three electrons the model can be viewed as a problem of a single electron in  three dimensions with impurity sites of strength $U_1$ and $U_2$ being placed on two infinite parallel lines, and this mapping can be extended for $N$ electrons. It seems that there will not be any rectification in this simplified version even beyond the two-particle scattering approximation and it can be confirmed by considering three-particle scattering explicitly.

\section{Discussion}To conclude we find that the IRLM with different tunnel junctions acts as a rectifier for  nonlinear dispersion of the leads while it can not rectify for linear dispersion. The IRLM is integrable by the scattering Bethe-Ansatz for linear dispersion, but the Bethe-Ansatz technique is not applicable for nonlinear dispersion. The current in Eq.(\ref{current}) is symmetric with respect to the bias as the distribution functions in Eq.(\ref{dist}) are independent of the tunneling junctions. Now the equations in Eq.(\ref{dist}) are valid only for the energy of the resonance level $(\epsilon_d)$ being greater than the energy of the upper bounds $(k^1_0~ {\rm and}~ k^2_0)$ on the distribution in momenta in both the leads. The limitation on the validity of Eq.(\ref{dist}) arises from the fact that the derivation of Eq.(\ref{dist}) does not include the bound-state contributions coming from the poles of the scattering matrix   \cite{Mehta07}. %In fact, it can be shown explicitly that the two-particle scattering state and the corresponding two-particle scattering matrix contain a two-particle bound state which behaves as an effective single composite particle in transport. 
In fact, it can be shown that the two-particle scattering states contain a two-particle bound state and the corresponding two-particle current is asymmetric for asymmetric junctions even for a linear dispersion \cite{Shen07}. In our study of lattice models using the LS  scattering approach \cite{Dhar08, RoySoori09}, we include the contributions from the two-particle bound states. %$\epsilon_d>k^1_0,~k^2_0$ is also a bit unphysical as most interesting phenomena, such as, a two-particle resonance \cite{RoySoori09} occur for the $\epsilon_d$ being within the chemical potential of the leads. 
Thus the symmetric transport in the IRLM with linear dispersion most probably arises only for the value of $\epsilon_d$ which does not include bound states. 
The scattering Bethe-Ansatz technique of Mehta-Andrei is incomplete as it does not capture the contributions from the bound states. 

The Bethe-Ansatz momenta $\{p_j\}$ in the scattering Bethe-Ansatz method are determined using periodic boundary conditions in an auxiliary algebraic Bethe-Ansatz problem and it has been claimed that for infinite periodicity these momenta will coincide with those of the physical systems. But it is not clear what is the mechanism of dissipation (or exchange of energy) of the scattered particles in the leads within such an approach. On the contrary, in the technique used here for the lattice model, it has been assumed as the original Landauer-B{\"u}ttiker (LB) scattering approach that all the dissipation occurs in the reservoirs connected to the leads. One can show an one-to-one connection between the LS scattering theory and the LB scattering approach. The present technique based on the LS scattering theory has been successfully applied to study transport in different interacting mesoscopic systems \cite{Dhar08, RoySoori09, Roy09}. Also the many-particle current within this technique merges with the current derived from the LB approach for noninteracting systems \cite{Dhar08}. Though there are a few numerical studies such as Ref.\cite{Boulat08} using the density matrix renormalization group (DMRG), to our best knowledge this is the first analytical study of nonlinear transport in symmetric/asymmetric IRLM for nonlinear dispersion. 

Recently it has been shown in Ref.\cite{Segal08} that thermal transport is symmetric/asymmetric with respect to temperature difference in mono-mode mediated energy exchange  between two metals for the linear/nonlinear dispersion of the metals. But in that study the two metallic leads are coupled by a single harmonic modes, i.e., linear interaction. Here we emphasize that in the IRLM due to local nonlinear interaction particles can exchange energy after scattering. Thus we expect to find rectification in the IRLM with asymmetric junctions for both the linear and the nonlinear dispersion of the leads. Finally, we can understand physically the mechanism behind rectification in the lattice model with asymmetric tunneling. Let we consider scattering of two electrons from one lead to another mediated by nonlinear interactions at impurity site. Two electrons with incident momenta $\{k_1,k_2\}$ can scatter into different channels with momenta $\{k'_1,k'_2\}$ satisfying the total energy conservation. The asymmetry in tunnel junctions creates a difference in the redistribution of momenta after scattering from the interactions for electrons coming from the left or the right leads. Thus we find the two-particle current asymmetry with asymmetric junctions. For many particles if a finite bias is applied across the impurity then the asymmetry in the redistribution of momenta persists and that generates the asymmetry in current.

\section{acknowledgments}

The author would  like to express his thanks to N. Andrei, G. Palacios, D. Segal, and specially D. Sen for many useful discussions. The hospitality at ICTP, Trieste and IIS, the Unv. of Tokyo is gratefully acknowledged. The work has been partially funded by the DOE under Grant No. DE-FG02-05ER46204.

%\bibliography{refs}

\begin{thebibliography}{34}
\expandafter\ifx\csname natexlab\endcsname\relax\def\natexlab#1{#1}\fi
\expandafter\ifx\csname bibnamefont\endcsname\relax
\def\bibnamefont#1{#1}\fi
\expandafter\ifx\csname bibfnamefont\endcsname\relax
\def\bibfnamefont#1{#1}\fi
\expandafter\ifx\csname citenamefont\endcsname\relax
\def\citenamefont#1{#1}\fi
\expandafter\ifx\csname url\endcsname\relax
\def\url#1{\texttt{#1}}\fi
\expandafter\ifx\csname urlprefix\endcsname\relax\def\urlprefix{URL }\fi
\providecommand{\bibinfo}[2]{#2}
\providecommand{\eprint}[2][]{\url{#2}}

\bibitem{Aviram74} A. Aviram and M. A. Ratner, Chem. Phys. Lett. {\bf 29}, 277 (1974).

\bibitem{Geddes90} N. J. Geddes, D. J. Sandman, J. R. Sambles, D. J. Jarvis, and W. G. Parker, Appl. Phys. Lett. {\bf 56}, 1916 (1990); A. S. Martin, J. R. Sambles, and G. J. Ashwell, Phys. Rev. Lett. {\bf 70}, 218 (1993); C. Joachim, J. K. Gimzewski, and A. Aviram, Nature {\bf 408}, 541 (2000).  

\bibitem{Zhao05} J. Zhao, C. Zeng, X. Cheng, K. Wang, G. Wang, J. Yang, J. G.
Hou, and Q. Zhu, Phys. Rev. Lett. {\bf 95}, 045502 (2005).

\bibitem{Song98} A. M. Song, A. Lorke, A. Kriele, J. P. Kotthaus, W. Wegscheider, and M. Bichler, Phys. Rev. Lett. {\bf 80}, 3831 (1998); S. de Haan, A. Lorke, J. P. Kotthaus, W. Wegscheider, and M. Bichler,  {\it ibid.} {\bf 92}, 056806 (2004).

%\bibitem{Song98} A. M. Song, A. Lorke, A. Kriele, J. P. Kotthaus, W. Wegscheider, and M. Bichler, Phys. Rev. Lett. {\bf 80}, 3831 (1998).

%\bibitem{Hann04} S. de Haan, A. Lorke, J. P. Kotthaus, W. Wegscheider, and M. Bichler, Phys. Rev. Lett. {\bf 92}, 056806 (2004).

\bibitem{Stopa02} M. Stopa, Phys. Rev. Lett. {\bf 88}, 146802 (2002); A. Vidan, R. M. Westervelt, M. Stopa, M. Hanson, and A. C. Gossard, Appl. Phys. Lett. {\bf 85}, 3602 (2004).

\bibitem{Ono02} K. Ono, D. G. Austing, Y. Tokura, and S. Tarucha, Science {\bf 297}, 1313 (2002).

\bibitem{Datta97} S. Datta, W. Tian, S. Hong, R. Reifenberger, J. I. Henderson, and C. P. Kubiak, Phys. Rev. Lett. {\bf 79}, 2530 (1997).

\bibitem{Mujica02} V. Mujica, M. A. Ratner, and A. Nitzan, Chem. Phys. {\bf 281}, 147 (2002).

\bibitem{Kornilovitch02} P.E. Kornilovitch, A. M. Bratkovsky, and R. Stanley Williams, Phys. Rev. B {\bf 66}, 165436 (2002).

\bibitem{Stokbro03} K. Stokbro, J. Taylor, and M. Brandbyge, J. Am. Chem. Soc. {\bf 125}, 3674 (2003). 

%\bibitem{Kornilovitch02} P.E. Kornilovitch, A. M. Bratkovsky, and R. Stanley Williams, Phys. Rev. B {\bf 66}, 165436 (2002).

\bibitem{Wu09} Recently sufficient conditions for thermal rectification have been identified; see, L-A. Wu and D. Segal, Phys. Rev. Lett. {\bf 102}, 095503 (2009).

\bibitem{Filyov80} V. M. Filyov, A. M. Tsvelick, and P. B. Wiegmann, Phys. Lett. {\bf 81A}, 175 (1980); P. Schlottmann, Phys. Rev. B {\bf 25}, 4815 (1982).

\bibitem{Mehta06} P. Mehta and N. Andrei, Phys. Rev. Lett. {\bf 96}, 216802 (2006).

%\bibitem{Mehta07} P.Mehta, S-P. Chao, and N. Andrei, Cond-mat/0703426. 

\bibitem{Doyon07} B. Doyon, Phys. Rev. Lett. {\bf 99}, 076806 (2007).

\bibitem{Borda08} L. Borda, A. Schiller, and A. Zawadowski, Phys. Rev. B {\bf 78}, 201301(R), (2008).

\bibitem{Boulat08} E. Boulat, H. Saleur, and P. Schmitteckert, Phys. Rev. Lett. 
{\bf 101}, 140601 (2008).

\bibitem{Wingreen94} N. S. Wingreen and Y. Meir, Phys. Rev. B {\bf 49}, 11040 
(1994).

\bibitem{Schiller95} A. Schiller and S. Hershfield, Phys. Rev. B {\bf 51}, 12896(R), (1995).
%\bibitem{Mehta06} P. Mehta and N. Andrei, Phys. Rev. Lett. {\bf 96}, 216802 (2006); P.Mehta, S-P. Chao, and N. Andrei, Cond-mat/0703426. 
\bibitem{Dhar08} A. Dhar, D. Sen, and D. Roy, Phys. Rev. Lett. {\bf 101}, 066805 (2008).

%\bibitem{Boulat08} E. Boulat, H. Saleur, and P. Schmitteckert, Phys. Rev. Lett. {\bf 101}, 140601 (2008).

\bibitem{RoySoori09} D. Roy, A. Soori, D. Sen, and A. Dhar,  Phys. Rev. B {\bf 80}, 075302 (2009).
\bibitem{Ralph94} D. C. Ralph and R. A. Buhrman, Phys. Rev. Lett. {\bf 72}, 3401 (1994); D. Goldhaber-Gordon, H. Shtrikman, D. Mahalu, D. Abusch-Magler, U. Meirav, and M. A. Kastner, Nature {\bf 391}, 156 (1998); R. M. Potok, I. G. Rau, H. Shtrikman, Y. Oreg, and D. Goldhaber-Gordon, Nature {\bf 446}, 167 (2007).

\bibitem{Mehta07} P.Mehta, S-P. Chao, and N. Andrei, Cond-mat/0703426 (unpublished). 

\bibitem{Morita71} T. Morita, J. Math. Phys. (N.Y.) {\bf 12}, 1744 (1971).

\bibitem{Shen07} J.-T. Shen and S. Fan, Phys. Rev. Lett. {\bf 98}, 153003 (2007); A. Nishino, T. Imamura, and N. Hatano, Phys. Rev. Lett. {\bf 102}, 146803 (2009).

\bibitem{Roy09} D. Roy, Phys. Rev. B {\bf 80}, 245304 (2009).

\bibitem{Segal08} D. Segal, Phys. Rev. Lett. {\bf 100}, 105901 (2008).


\end{thebibliography}

\end{document}